\documentstyle[aps,preprint]{revtex}
\begin{document}
\draft

\title{Static properties of a quasi one-dimensional
antiferromagnet in magnetic field}

\author{M.~E.~Zhitomirsky}
\address{Institute for Solid State Physics, University of Tokyo,
Tokyo 106, Japan \\
and L.~D.~Landau Institute for Theoretical Physics, Moscow 117334
Russia}
\author{I.~A.~Zaliznyak}
\address{P.~L.~Kapitza Institute for Physical Problems, Moscow 117334,
Russia \\
and DRFMC--Centre d'Etudes Nucl\'eaires de Grenoble,
BP 85X, 38041 Grenoble C\'edex, France}

\date{11 October, 1995}

\maketitle

\begin{abstract}
We calculate the effect of zero-point fluctuations on the
magnetostatic properties of a quasi one-dimensional
antiferromagnetic helimagnet with and without in-plane anisotropy.
We find substantial reduction of the uniform susceptibility and
the sublattice magnetization from their classical values and
appreciable renormalization of the spin-flop field at $T=0$. The
magnetization curve varies nonlinearly at small fields and is
described by a universal formula at high fields. The results are
compared with numerical simulations on one-dimensional systems and
measurements in the CsNiCl$_3$-type compounds.
\end{abstract}

\newpage

\section{Introduction}

Recently, there has been growing interest in the behavior of
one-dimensional and quasi one-dimensional antiferromagnets in
strong magnetic fields. In the former case much attention was
devoted to the destruction of the Haldane state of a spin-1
antiferromagnetic chain in applied field and to the magnetization
process at higher fields
\cite{NENP,Ajiro,Bethe,Affleck,S=1,crit,T>0}. In the latter
case the interest lies mainly in properties of CsNiCl$_3$-type
magnetic compounds which are good model Heisenberg systems. They
contain a stacked triangular lattice of antiferromagnetically
coupled $3d$ ions with the intrachain exchange constant $J$ being
several orders of magnitude larger than the interchain coupling
$J'$. Because of such smallness of interchain interactions,
quantum effects play an important role in the N\'eel phase.
Extensive theoretical works have been presented to describe
quantum renormalizations of the magnon spectrum at $T=0$ in these
materials \cite{QR1,QR2,Shiba}. Though experiments indicate
significant deviations in static properties of quasi
one-dimensional antiferromagnets from predictions of the classical
theory as well \cite{Yelon-Cox,Eibshutz,AZ,Zaliznyak,AP,Luethi,Goto},
only a few effects were considered theoretically
\cite{Shiba,Weltz,Watabe}.

Here we study the lowest order quantum corrections to the static
characteristics of a quasi one-dimensional antiferromagnetic
helimagnet: staggered magnetization, transverse and in-plane
magnetic susceptibilities, spin-flop field, and full magnetization
curve. We show that the magnetization process at high fields is
the same for all low-dimensional systems, i.e., is independent
from the nature of the ground state at $H=0$. This fact is
connected with the different corrections to the staggered and
uniform magnetizations appearing in the spin-wave theory.

The model is defined by the Heisenberg Hamiltonian for a system of
$N$ equivalent spins on a simple Bravais lattice with a
single-ion anisotropy and the Zeeman term:
\begin{equation}
\hat{\cal H} = \sum_{i,j} J_{ij}\,
{\bf S}_i \cdot {\bf S}_j + D\sum_i(S_i^z)^2
- \sum_i \gamma {\bf H}\cdot {\bf S}_i\ ,
\label{H}
\end{equation}
(in the following, we include the gyromagnetic ratio
$\gamma=g\mu_B$ in the definition of $H$). Without anisotropy and
magnetic field the classical ground state of (\ref{H}) is a plane
spin helix ${\bf S}_i = (S\cos \theta_i, S\sin\theta_i,0)$,
$\theta_i= {\bf Q\cdot r}_i$, where the ordering wave vector $\bf
Q$ corresponds to the minimum of the Fourier transform of exchange
interactions: $J_{\bf k} = \sum_i J_i \exp(i{\bf k\cdot r}_i)$
\cite{Nagamiya}.

To obtain the first $1/S$ correction to the classical energy of
the ordered state we use the standard perturbation scheme in which
spin operators are replaced by their bosonic equivalents by
applying the reduced Holstein-Primakoff transformation
$S^z=S-a^+a$, $S^+=\sqrt{2S}a$, $S^-=\sqrt{2S}a^+$ to each spin in
its local coordinate system. The first order corrections appear in
this harmonic approximation due to the quadratic part of the boson
Hamiltonian which corresponds to the quantized classical spin
waves with zero-point oscillations. Diagonalization of the boson
Hamiltonian is trivial for a helimagnet without in-plane
anisotropy ($D>0$, ${\bf H}\parallel\hat{\bf z}$). Analytical
expressions for quantum effects under these conditions are
given in Sec.~II. Less studied in the literature case of a
bunched spiral in the presence of an in-plane anisotropy is
considered in Sec.~III. Treating perturbatively helix distortions
we calculate the classical spin-wave spectrum and find the quantum
effects for a spin helix with the easy-axis anisotropy and
in-plane magnetic field. Comparison of the theory with
experimental measurements in CsNiCl$_3$ and CsMnBr$_3$ is
presented in Sec.~IV.

\section{Axially symmetric case}

We first present standard spin-wave calculations
for a helimagnet with the
easy-plane type of anisotropy ($D>0$) and magnetic field parallel
to the symmetry axis $\hat{\bf z}$ (see, e.g., Ref.~\cite{Nagamiya})
and find analytical expressions for quantum effects in the quasi
one-dimensional limit. For this axially symmetric case spins form
an umbrella structure ${\bf S}_i = (S \cos\alpha \cos\theta_i, S
\cos\alpha \sin\theta_i, S\sin\alpha)$, where $\theta_i= {\bf
Q\cdot r}_i$. By minimizing the classical energy of the deformed
spin helix one obtains an expression for the canting angle
$\sin\alpha = H/H_s$, which is valid up to the saturation field
$H_s=2(J_0-J_{\bf Q}+D)S$. Above this critical field
$\sin\alpha=1$ and all spins are aligned parallel to $\bf H$.

The quadratic part of the Hamiltonian after the normal ordering
and Fourier transformation of boson operators is written in the
form:
\widetext
\begin{equation}
\hat{\cal H}_2 = \case{1}{2} NDS\cos^2\alpha + \sum_{\bf k}
\left[ (A_{\bf k}+C_{\bf k})a^+_{\bf k}a^{_{}}_{\bf k}
+ \case{1}{2} B_{\bf k}(a^{_{}}_{\bf k}a^{_{}}_{\bf -k} +
a_{\bf k}^+a_{\bf -k}^+)\right],
\label{H2}
\end{equation}
with
\begin{eqnarray*}
A_{\bf k} & = & -2SJ_{\bf Q} + S\cos^2\!\alpha (J_{\bf k} + D)
+ \case{1}{2} S(1+\sin^2\!\alpha)\,(J_{\bf k+Q}
+ J_{\bf k-Q})\  , \\
B_{\bf k} & = & -S\cos^2\!\alpha \left[ J_{\bf k}+D -
\case{1}{2} (J_{\bf k+Q}+J_{\bf k-Q})\right] , \quad
C_{\bf k}  =  S\sin\alpha(J_{\bf k+Q}-J_{\bf k-Q}) \ .
\end{eqnarray*}
\narrowtext
Note, that we work in the extended zone scheme assigning
wave vectors to the paramagnetic Brillouin zone.
Coefficients $A_{\bf k}$ and $B_{\bf k}$ are even functions of
$\bf k$, while $C_{\bf k}$ is an odd function. Performing the
standard Bogolyubov transformation, which leaves odd-$\bf k$ terms
unchanged, one can diagonalize the bilinear form of boson
operators:
\begin{equation}
\sum_{\bf k}\left[ \omega_{\bf k}\left(b_{\bf k}^+b_{\bf k}^{_{}}
+ \case{1}{2} \right) - \case{1}{2} A_{\bf k}\right] ,
\label{GSEP}
\end{equation}
where the classical spin-wave frequencies are given by
\begin{equation}
\omega_{\bf k} = \sqrt{A_{\bf k}^2-B_{\bf k}^2} + C_{\bf k} \ .
\label{swp}
\end{equation}
The easy-plane anisotropy creates gaps at ${\bf k}=\pm{\bf Q}$,
whereas spin waves at ${\bf k}\rightarrow0$ remain to be
the Goldstone modes up to the saturation field.

We calculate the magnetization at $T=0$ taking the derivative of
the ground state energy with respect to the magnetic field
\widetext
\begin{equation}
M = N\frac{H}{2(J_0-J_{\bf Q}+D)}\left[1-\frac{1}{2SN}\sum_{\bf k}
\frac{\frac{1}{2}(J_{\bf k+Q}+J_{\bf k-Q})-J_{\bf k}-D}
{J_0-J_{\bf Q}+D}
\sqrt{\frac{A_{\bf k}+B_{\bf k}}{A_{\bf k}-B_{\bf k}}}
\right] .
\label{M1}
\end{equation}
\narrowtext
The factor before brackets is the magnetization value in the
classical theory which is renormalized due to zero-point
oscillations. The expression (\ref{M1}) corresponds to the field
region $0<H<H_s$, whereas in the completely aligned phase above
$H_s$ quantum corrections disappear and magnetization saturates:
$M_s=NS$. The average value of the local spin below $H_s$ is given
by the standard formula \cite{Nagamiya}
\begin{equation}
S_{\rm av} = S\left[1-\frac{1}{2SN}\sum_{\bf k}
\left(\frac{A_{\bf k}} {\sqrt{A_{\bf k}^2-B_{\bf k}^2}}
- 1\right)\right] .
\label{<S>1}
\end{equation}

In the general case correction terms in (\ref{M1}) and
(\ref{<S>1}) should be evaluated by numerical integration over the
three-dimensional Brillouin zone. However, analytical expressions
representing the leading logarithmic terms can be found in the case
of quasi one-dimensional magnets. Let us calculate them for
hexagonal CsNiCl$_3$-type antiferromagnets. Using the rectangular
coordinate system, it is convenient to measure wave-vector
components along the $z$ and $x$ axes in units of the
reciprocal intersite distances $c$ and $a$ respectively, while the
component in the $y$ direction in units of $2/\sqrt{3}a$. Then,
the Fourier transform of the exchange energy takes the form
\begin{equation}
J_{\bf k}  =  2J\cos k_z + 6 J' \gamma_{\bf k} \ , \quad
\gamma_{\bf k} = \case{1}{3} (\cos k_x +
2 \cos \case{1}{2} k_x\cos k_y) \ ,
\label{FT}
\end{equation}
with each wave-vector component varying between $0$ and $2\pi$.
The helix vector of the triangular structure found by minimization
of $J_{\bf k}$ is ${\bf Q} = (4\pi/3,0,\pi)$ in this coordinate
frame.

Keeping only the leading contributions in the quasi
one-dimensional limit $J\gg J',D$, we omit terms proportional to
$J'$ and $D$ in numerators of Eqs.~(\ref{M1}) and (\ref{<S>1}) and
otbtain the following expressions ($H_s\approx8JS$):
\widetext
\begin{equation}
m=M/N = \frac{H}{8J}\left[ 1 - \frac{1}{2S}\int\frac{d^3 k}{(2\pi)^3}
\frac{(-\cos k_z)\sqrt{1-\cos k_z}}{\sqrt{1+\cos k_z + p_{\bf k}
-H^2/(32J^2S^2)\cos k_z}}\right] ,
\label{M2}
\end{equation}
\begin{equation}
S_{\rm av} = S\left[ 1 - \frac{1}{2S}\int\frac{d^3 k}{(2\pi)^3}
\left(\frac{1-H^2/(64J^2S^2)\cos k_z}
{\sqrt{(1-\cos k_z+q_{\bf k})(1+\cos k_z +p_{\bf k}
-H^2/(32J^2S^2)\cos k_z)}}-1\right)\right] ,
\label{<S>2}
\end{equation}
where
$$
p_{\bf k}=\frac{3J'}{2J}(1+2\gamma_{\bf k})+\frac{D}{2J} \ , \quad
q_{\bf k} = \frac{3J'}{2J}(1-\gamma_{\bf k}) \ .
$$
The integration over $k_z$ in (\ref{M2}) can be done explicitly
for arbitrary $p_{\bf k}$ and $H$. However, this would be
an overestimation as we have already neglected linear and higher
order terms in $J'/J$. Therefore at $H\ll H_s$ we integrate
(\ref{M2}) keeping only logarithmic and zero order in $p_{\bf k}$
terms. This yields:
\begin{equation}
m = \frac{H}{8J}\left\{1 - \frac{1}{2\pi S} \left[
\ln\frac{16J}{3J'} - 2 - \langle\ln(1+2\gamma_{\bf k} +
D/3J' + H^2/48JJ'S^2)\rangle\right]\right\},
\label{M3}
\end{equation}
where angular brackets denote averaging over the two-dimensional
Brillouin zone. The sublattice magnetization is given with the
same accuracy $O[(J'/J)\ln(J/J')]$ by
\begin{equation}
 S_{\rm av} = S\left\{1 - \frac{1}{2\pi S}
\left[\ln\frac{16J}{3J'} - \pi -
\case{1}{2} \langle\ln(1 - \gamma_{\bf k})\rangle -
\case{1}{2} \langle\ln(1+2\gamma_{\bf k}+D/3J' +
H^2/48JJ'S^2)\rangle\right]\right\}.
\label{<S>3}
\end{equation}
\narrowtext
The main part of quantum effects for both $m$ and $S_{\rm av}$ is
represented at $H=0$ by the large logarithm $\ln(16J/3J')$,
which is in agreement with a qualitative suggestion of
Ref.~\cite{Zaliznyak}. Numerical constants yield in the isotropic
case only small corrections:
$$
\langle \ln(1+2\gamma_{\bf k})\rangle =  -0.452 \ ,\quad
\langle\ln(1-\gamma_{\bf k})\rangle  =  -0.176 \ .
$$
At low fields the magnetization is a linear function of
$H$ with the susceptibility per site
\begin{equation}
\chi_\perp = \frac{1}{8J}\left\{ 1- \frac{1}{2\pi S}\left[
\ln\frac{16J}{3J'} - 2 - \langle\ln(1+2\gamma_{\bf k} +
D/3J')\rangle\right]\right\}.
\label{chip}
\end{equation}
The additional correction terms in (\ref{M3}) and (\ref{<S>3}),
however, grow with increasing field and give nonlinear
dependences of magnetization and spin length. This nonlinear
increase of $m$ and $S_{\rm av}$ means suppression of quantum
fluctuations in magnetic field, which must completely disappear in
the collinear phase at $H\ge H_s$.

In the isotropic case Eqs.~(\ref{<S>3}) and (\ref{chip}) give good
coincidence up to 2--3\% with the results by Ohyama and Shiba
\cite{Shiba}, who estimated integrals in (\ref{M1}) and
(\ref{<S>1}) numerically. Note, that all the derived expressions
are also valid for the easy-axis helimagnet at fields above the
spin-flop field (${\bf H}\parallel\hat{\bf z}$), when spins lie in
the basal plane and the axial symmetry is restored. In this case
the constant $D$ in (\ref{M3}) and (\ref{<S>3}) has the negative
sign as in the Hamiltonin (\ref{H}).

At high fields the magnitude of quantum corrections to $m$ becomes
independent of the interchain coupling. Crossover between two
regimes takes place at a characteristic field
$H_c=\sqrt{48J'JS^2}$. At $H_c\ll H \le H_s$ one can neglect the
term $p_{\bf k}$ in (\ref{M2}) and obtain after integration over
$k_z$ the magnetization
\widetext
\begin{eqnarray}
m & = & S\,\frac{H}{H_s} \left[ 1 - \frac{1}{\pi S}
\frac{H_s\sqrt{H_s^2-H^2}}{2H^2-H_s^2} \left( 1 -
\sqrt{\frac{H_s^2-H^2}{|2H^2-H_s^2|}}f(H)\right)\right], \label{M4}\\
& & {\rm where}\ \ f(H)=\left\{\begin{array}{ll}
     \ln\left(\sqrt{\frac{\displaystyle H_s^2}{\displaystyle H^2}-2}+
     \sqrt{\frac{\displaystyle H_s^2}{\displaystyle H^2}-1}\right) ,
    & H\le H_s/\sqrt{2}\ , \\
     \arcsin\sqrt{2-\frac{\displaystyle H_s^2}{\displaystyle H^2}}\ ,
    & H_s/\sqrt{2}\le H \le H_s\ .
     \end{array} \right.   \nonumber
\end{eqnarray}
\narrowtext
As for the staggered
magnetization $S_{\rm av}$, it remains finite in the field region
$0<H<H_s$ only if $J'\ne 0$.

According to (\ref{M4}) the magnetization approaches its
saturation value with the asymptotic behavior
$M = M_s[1 - \sqrt{2(1-H/H_s)}/\pi S]$. As we neglected the
interchain coupling, this result should be also valid for a single
antiferromagnetic chain. Hodgson and Parkinson \cite{Bethe} using
the Bethe ansatz approach found a different critical behavior for
the chain magnetization:
\begin{equation}
M = M_s\left[ 1 - 2\sqrt{1-H/H_s}/\pi S \right] .
\label{M7}
\end{equation}

The reason of this discrepancy can be explained as follows. In the
bosonic representation the Hamiltonian (\ref{H}) is equivalent to
a gas of Bose particles interacting via a $\delta$-potential.
Their interaction proportional to $1/S$ is neglected in the
harmonic approximation. As magnetic field approaches the
saturation field, the boson density tends to zero. In the
low-density limit the average kinetic energy of bosons is small
and they behave as if the potential has a hard core. Therefore, in
one dimension at fields sufficiently close to $H_s$ the
interaction between spin waves becomes non-negligible for any
value of $S$ and gives the modified asymptotic behavior
\cite{crit}. The width of the critical region of quantum
fluctuations decreases with increasing $S$.

We compare our expression for the magnetization (\ref{M4}) with
the results of exact numerical diagonalization of the Hamiltonian
for finite systems. As can be seen from Fig.~1, the formula
(\ref{M4}) is in remarkable agreement with the magnetization curve
for a spin-1 antiferromagnetic chain found numerically
by Sakai and Takahashi \cite{S=1} except for a
low-field region where the nonmagnetic Haldane phase is developed
and a close vicinity of $H_s$. Thus, we conclude that at the
intermediate field region the magnetic field suppresses
zero-point fluctuations and increases the accuracy of the
spin-wave theory. The magnetization curve for a system of coupled
antiferromagnetic chains does not depend at high fields on the
nature of the ground state (which is determined by $J'$ and $S$)
and is given by (\ref{M4}). However, in the narrow field region
near $H_s$ the interaction between spin-wave modes becomes
important and changes the asymptotic behavior of $M(H)$. The size
of this region is small already for $S=1$ and the harmonic
approximation of the spin-wave theory gives good account for most
of the experimental results at high fields \cite{NENP,Ajiro,Goto}.

\section{In-plane anisotropy}

Consider further quantum effects for a Heisenberg helimagnet with
an in-plane anisotropy. The easy-axis anisotropy or magnetic field
applied in the spin plane perturb spin helix adding higher-order
harmonics $n{\bf Q}$ \cite{Nagamiya,Z-Zh}. Derivations of quantum
corrections are similar for all cases. We present in detail
calculations for $D<0$ and $H=0$ listing in other cases only final
expressions.

If the anisotropy constant $D$ in (\ref{H}) is
negative, spins in the spiral structure rotate in the vertical
plane: ${\bf S}_i = (S\sin\theta_i, 0, S\cos\theta_i)$. As
the exchange interactions are much stronger than the relativistic
anisotropy we can consider only the first anharmonic term
\begin{equation}
\theta_i = {\bf Q \cdot r}_i - \varphi\sin (2{\bf Q \cdot r}_i) \ .
\label{dhel}
\end{equation}
Making substitution of (\ref{dhel}) into (\ref{H}), we find the
following expression for the deviation angle:
$$
\varphi = |D|/(J_{3\bf Q} - J_{\bf Q})\ .
$$
The formula (\ref{dhel}) is valid for both incommensurate and
commensurate antiferromagnets (if only $3{\bf Q}\ne\pm{\bf Q}$).
In the latter case there should be an additional constant term in
(\ref{dhel}) which describes lock-in of a spin helix with respect
to the crystal axes. However, we can neglect it, because the
commensuration energy appears only at the $N/2$ order of the
expansion in $|D|$, where $N {\bf Q} = {\bf G}$, ${\bf G}$ being
the reciprocal lattice vector \cite{Z-Zh}.

Taking into account (\ref{dhel}) the quadratic part of the
Hamiltonian is written as
\begin{eqnarray}
\hat{\cal H} & = & \mbox{} - \case{1}{4} N|D|S+
\sum_{\bf k} \left[ A_{\bf k}\,a^+_{\bf k}a^{_{}}_{\bf k}
+ \case{1}{2} B_{\bf k}(a^{_{}}_{\bf k}a^{_{}}_{\bf -k} +
a_{\bf k}^+a_{\bf -k}^+) \right. \nonumber \\
  &  & \left. \mbox{}+
 C_{\bf k}(a^+_{\bf k+Q}
a^{_{}}_{\bf k-Q} + {\rm c.c.}) +
\case{1}{2} D_{\bf k}(a^{_{}}_{\bf k+Q}a^{_{}}_{-\bf k+Q} +
a^{_{}}_{\bf k-Q}a^{_{}}_{\bf -k-Q} + {\rm c.c.})\right],
\label{H-eas1}
\end{eqnarray}
where
\begin{eqnarray*}
A_{\bf k} & = & -2J_{\bf Q}S + S\left[J_{\bf k} +
\case{1}{2} (J_{\bf k+Q} + J_{\bf k-Q})\right] + \case{1}{2} |D|S\ , \\
B_{\bf k} & = & -S\left[J_{\bf k} - \case{1}{2} (J_{\bf k+Q}+J_{\bf k-Q})
\right] - \case{1}{2} |D|S \ , \\
C_{\bf k} & = & \case{5}{4} |D|S +
\case{1}{2}\varphi S\left[J_{\bf k} - \case{1}{2}
(J_{{\bf k}+2{\bf Q}}+J_{{\bf k}-2{\bf Q}})\right], \quad
D_{\bf k}  =  C_{\bf k} - |D|S \ .
\end{eqnarray*}

As was shown by Elliot and Lange using general symmetry arguments
\cite{EL}, a helimagnet with in-plane magnetic field always have a
zero-frequency mode for spin rotations inside helix plane. Their
theorem fulfills obviously for another types of in-plane
anisotropy.  We illustrate this by finding
in the first order in $D$ analytical expressions
for the classical spectrum and the ground state energy of an
easy-axis helimagnet.

The Bogolyubov transformation can be used to diagonalize the first
two terms in (\ref{H-eas1}). After this procedure the boson part
of the Hamiltonian (\ref{H-eas1}) is reduced to
\begin{eqnarray}
\hat{\cal H} & = & \sum_{\bf k}\left[\omega^0_{\bf k}
\left(b_{\bf k}^+b_{\bf k}^{_{}} + \case{1}{2} \right) -
\case{1}{2} A_{\bf k} + \tilde{C}_{\bf k}
(b^+_{\bf k+Q} b^{_{}}_{\bf k-Q} + {\rm c.c.}) \right.
\nonumber \\
& & \left. \mbox{} + \case{1}{2} \tilde{D}_{\bf k}
(b^{_{}}_{\bf k+Q}b^{_{}}_{\bf -k+Q} +
b^{_{}}_{\bf k-Q}b^{_{}}_{\bf -k-Q} + {\rm c.c.}) \right],
\label{H-eas2}
\end{eqnarray}
where
$$
\begin{array}{ccl}
\omega^0_{\bf k} & = & \sqrt{A_{\bf k}^2 - B_{\bf k}^2}\ ,
\quad \tanh 2\psi_{\bf k} = -B_{\bf k}/A_{\bf k} \ , \\
\tilde{C}_{\bf k} & = & C_{\bf k}\cosh(\psi_{\bf k+Q}+\psi_{\bf k-Q}) +
D_{\bf k}\sinh(\psi_{\bf k+Q}+\psi_{\bf k-Q})\ , \\
\tilde{D}_{\bf k} & = & C_{\bf k}\sinh(\psi_{\bf k+Q} +
\psi_{\bf k-Q}) + D_{\bf k} \cosh(\psi_{\bf k+Q}+\psi_{\bf k-Q}) \ .
\end{array}
$$
Note, that constants $\omega^0_{\bf k}$, $A_{\bf k}$ and
$B_{\bf k}$ are given by the same expressions as in the case of
the easy-plane anisotropy (\ref{H2}) but with $D_{\rm eff} = |D|/2$.
Attempt to find eigenfrequencies of (\ref{H-eas2}) leads to an
infinite system of coupled equations. However, the diagonalization
may be simplified since the amplitudes $\tilde{C}_{\bf k}$,
$\tilde{D}_{\bf k}$ are of the order of $|D|S$ everywhere in the
Brillouin zone except for vicinity of the point ${\bf k} =
0$, where the Bogolyubov transformation becomes nonanalytical
in the limit $|D|\rightarrow0$. Therefore umklapp terms in the
Hamiltonian (\ref{H-eas2}) are important for the calculation of the
spin-wave spectrum in the first order in $|D|S$ only near ${\bf
k}=\pm {\bf Q}$, where they describe interaction of four nearly
degenerate magnons with propagation vectors ${\bf \pm Q\pm q}$.

In the following calculations we use the method proposed by
White \cite{White}. We define the column vector of bose operators
$\hat{x}_{\bf q}=(b^{_{}}_{\bf q+Q},b^{_{}}_{\bf q-Q},
b_{\bf-q+Q}^+,b_{\bf-q-Q}^+)$.
Then, excluding the constant term $-\frac{1}{2}\sum_{\bf k}A_{\bf k}$,
the Hamiltonian (\ref{H-eas2}) can be rewritten as
\begin{equation}
\hat{\cal H} = \sum_{\bf q} \hat{x}^+_{\bf q}{\cal H}^{_{}}_{\bf q}
\hat{x}^{_{}}_{\bf q}\ ,
\end{equation}
where
\begin{equation}
{\cal H}_{\bf q} = \left(\begin{array}{cccc}
\omega^0_{\bf q+Q} &\tilde{C}_{\bf q} &0 & \tilde{D}_{\bf q}\\[1mm]
\tilde{C}_{\bf q} &\omega^0_{\bf q-Q} &\tilde{D}_{\bf q} &
0\\[1mm]
0&\tilde{D}_{\bf q}&\omega^0_{\bf q+Q} &\tilde{C}_{\bf q}\\[1mm]
\tilde{D}_{\bf q} & 0 & \tilde{C}_{\bf q} & \omega^0_{\bf q-Q}
\end{array}\right) \ .
\label{4na4}
\end{equation}
Following the above arguments, we replace the eigenvalue problem
for the infinite matrix $\hat{\cal H}$ by diagonalization of $4\times4$
matrix (\ref{4na4}). The transformation
$\hat{x}_{\bf k}=S\hat{y}_{\bf k}$ that diagonalizes
${\cal H}_{\bf q}$ must preserve the
commutator $[\hat{x}^{_{}}_{\bf k},\hat{x}_{\bf k}^+]=g$, where the
metric matrix $g$ has only diagonal elements: $(+1,+1,-1,-1)$.
Solving equation
\begin{equation}
S^{-1}g{\cal H}_{\bf q} S = g \Omega_H \ ,
\end{equation}
we obtain the diagonal matrix $\Omega_H$ with two pairs of equal
eigenfrequencies
\begin{eqnarray}
(\omega^\pm)^2  & = & \case{1}{2} [(\omega^0_{\bf q+Q}) +
(\omega^0_{\bf q-Q})^2] + \tilde{C}_{\bf q}^2 -
\tilde{D}_{\bf q}^2 \pm \left\{\case{1}{4}
[(\omega^0_{\bf q+Q})^2 - (\omega^0_{\bf q-Q})^2]^2
\right. \nonumber\\
& & \left. \mbox{} + \tilde{C}_{\bf q}^2\,
(\omega^0_{\bf q+Q} + \omega^0_{\bf q-Q})^2 - \tilde{D}_{\bf q}^2\,
(\omega^0_{\bf q+Q} - \omega^0_{\bf q-Q})^2 \right\}^{1/2} ,
\end{eqnarray}
which can be mapped onto two different magnon branches in a half of the
Brillouin zone. At $\bf k=Q$ the frequency gap of the first branch
is $\omega^+_{\bf Q}= 2S[|D|(J_0+J_{2\bf Q}-2J_{\bf Q})]^{1/2}$,
while $\omega^-_{\bf Q}=0$. Thus, the classical spectrum of a
helimagnet with the easy-axis anisotropy contains two quasi
Goldstone modes at ${\bf k}=0$ and $\bf Q$, where the additional
zero-gap mode results from the absence of the basal plane
anisotropy in the Hamiltonian (\ref{H}). We emphasize again that
this conclusion is valid for a commensurate helimagnet only in the
first order in $|D|$. For example, the second mode at $\bf k=Q$ in
triangular antiferromagnets acquires a frequency gap proportional
to $|D|^{3/2}$ \cite{Z-Zh}.

The energy of zero-point fluctuations is determined after the
last diagonalization by $E^{(0)} = \frac{1}{2} \sum_{\bf k > 0}
(\omega^+_{\bf k}+\omega^-_{\bf k})$. However, the umklapp terms
in (\ref{H-eas2}) yield correction to the energy of the ground
state only in the second order in anisotropy constant and,
therefore, should be omitted. The ground state
energy in the first order in $|D|$ can be obtained after partial
diagonalization:
\begin{equation}
E_{\rm g.s.} = J_{\bf Q}S(S+1) - \case{1}{2} |D|S(S+1) +
\case{1}{2} \sum_{\bf k} \omega^0_{\bf k} \ .
\label{GSE}
\end{equation}
To calculate the staggered magnetization with the same accuracy one
has, in contrast, to determine explicitly the transformation matrix $S$.
Expressing operators $a_i$ via new boson modes, it is possible
to find the average spin length including its modulation
[$\sim\cos(2{\bf Q}\cdot{\bf r}_i)$] in a bunched spiral.
These calculations cannot be done analytically and require
numerical study which was presented in the case of triangular
antiferromagnets by Watabe {\it et al\/}. \cite{Watabe}.
Note, that for a small anisotropy (the case
with which we are only concerned here)
the sublattice magnetization can be roughly estimated as in the
isotropic case by using Eq.~(\ref{<S>3}) with $D=0$.

Expanding
(\ref{GSE}) up to the first order in $D$, we get the
anisotropy energy of a triangular antiferromagnet renormalized  by
zero-point oscillations
\begin{equation}
E^{(1)}_{\rm an} =  - \frac{|D|S(S+1)}{2} + \frac{|D|S}{4\pi}
\left[\ln\frac{16J}{3J'} - \langle\ln(1 + 2\gamma_{\bf k})\rangle
\right].
\label{E1}
\end{equation}
If magnetic field is applied along the easy axis, the vertical
position of the helix plane becomes unstable at high fields.
Magnetic energy exceeds the relativistic anisotropy, and
above the spin-flop field $H_{\rm sf}$ spins lie in the basal plane.
In the classical picture, anisotropic part of the energy for this
configuration $E^{(2)}_{\rm an}$ vanishes. However, zero-point
motion of quantum spins makes $E^{(2)}_{\rm an}\ne 0$. Expanding
the ground state energy (\ref{GSEP}) in powers of $D$,
we have in the first order
\begin{equation}
E^{(2)}_{\rm an} =  - \frac{|D|S}{2\pi}
\left[\ln\frac{16J}{3J'} -  \langle\ln(1 +
2\gamma_{\bf k})\rangle\right].
\label{E2}
\end{equation}
The energy difference for vertical and horizontal configurations
of polarization plane $E_{\rm an} = E^{(2)}_{\rm an} -
E^{(1)}_{\rm an}$ will be used below for the calculation of the
spin-flop field.

Consider now the effect of magnetic field applied along the
easy axis. The distortion of a spin helix is described in this
case by
$$
\theta_i = {\bf Q\cdot r}_i - \alpha\sin({\bf Q \cdot r}_i) -
\varphi\sin(2{\bf Q \cdot r}_i)
$$
with $\alpha = H/(J_0 + J_{2\bf Q}-2J_{\bf Q})S$
\cite{Nagamiya,Z-Zh}. As we explained above, the
spin harmonics generated by single-ion anisotropy, being
important for calculations of the correct classical spin-wave
spectrum, can be neglected for the estimate of the ground state
energy in the first order in $D$. For the latter it is
necessary to take into account only changes in constants $A_{\bf
k}$ and $B_{\bf k}$. The new spin harmonic generated by magnetic
field gives
\begin{equation}
\delta A_{\bf k} = \delta B_{\bf k} = - \case{1}{4} \alpha^2 S
\left[J_{\bf k-Q} + J_{\bf k+Q} - J_{\bf k} - \case{1}{2}
(J_{{\bf k}-2{\bf Q}} + J_{{\bf k}+2{\bf Q}})\right].
\end{equation}
Magnetic field also generates umklapp terms in the boson
Hamiltonian with a momentum transfer $\pm\bf Q$. These terms
are proportional to $\alpha$ and, therefore, should be taken into
account for calculation of the ground state energy with the
accuracy $O(\alpha^2)\sim H^2$. Inspection of corresponding
expressions reveals, however, that umklapp terms have an
additional smallness of the order $J'/J$ in the quasi
one-dimensional limit. Having in mind this case, one can calculate
magnetic contribution to the ground state energy without umklapp
terms as in the above analysis of the easy-axis anisotropy. The
uniform magnetization found in this approximation is
\widetext
\begin{equation}
m  =  \frac{H}{2(J_0+J_{2\bf Q}-2J_{\bf Q})}
\left[1-\frac{1}{2SN} \sum_{\bf k}
\frac{J_{\bf k} + \frac{1}{2}
(J_{{\bf k}+2{\bf Q}} + J_{{\bf k}- 2{\bf Q}}) - J_{\bf k-Q} -
J_{\bf k+Q}} {J_0+J_{2\bf Q}-2J_{\bf Q}}
\sqrt{\frac{A_{\bf k}-B_{\bf k}}{A_{\bf k}+B_{\bf k}}} \right] .
\label{Mxz}
\end{equation}
Analytical estimate of the integral in (\ref{Mxz}) yields the
following expression for the in-plane susceptibility of a quasi
one-dimensional triangular antiferromagnet
\begin{equation}
\chi_\parallel = \frac{1}{16J} \left\{1 - \frac{1}{2\pi S}
\left[ \ln\frac{16J}{3J'} - 2 -
\langle\ln(1-\gamma_{\bf k})\rangle \right] \right\} .
\label{chi-par}
\end{equation}
Note, that quantum corrections break the relation $\chi_\perp =
\chi_\parallel$, which holds for a classical antiferromagnetic
helimagnet in the quasi one-dimensional limit.

With the help of (\ref{chip}), (\ref{E1}), (\ref{E2}), and
(\ref{chi-par}) we find the critical field of the
spin-flop transition as a field at which anisotropy and magnetic
energies of a quantum helimagnet are equal:
\begin{equation}
H^2_{\rm sf} = 16DJS^2 \frac{1 + 1/S - 3/(2\pi S)[\ln(16J/3J')
- \langle\ln(1 + 2\gamma_{\bf k})\rangle]}
{1 - 1/(2\pi S)[\ln(16J/3J') - 2 -
2\langle\ln(1 + 2\gamma_{\bf k})\rangle + \langle\ln(1 -
\gamma_{\bf k})\rangle]} \ .
\label{Hsf}
\end{equation}
\narrowtext

Let us list at the end the results for two another important
cases, which can be derived in a similar way: (i) $D<0$,
${\bf H}\perp\hat{\bf z}$, the magnetization is
given by the formula for the easy-plane anisotropy (\ref{M3}) with
the substitution $D\rightarrow|D|/2$; (ii) $D>0$, ${\bf H} \perp
\hat{\bf z}$, in-plane susceptibility in this case
coincides with (\ref{chi-par}).

\section{Discussion}

We compare our theoretical results with measurements on two
isostructural quasi one-dimensional hexagonal antiferromagnets
CsNiCl$_3$ and CsMnBr$_3$, for which a lot of experimental data
are available including the zero-point
spin reductions \cite{Yelon-Cox,Eibshutz} and precise magnetization
curves \cite{AZ,Zaliznyak}. Both compounds are considered to be
ideal Heisenberg systems described by the Hamiltonian (1)
and differ only in sign and relative strength of the anisotropy
constants.

First we consider the case of CsMnBr$_3$, which has a large spin
($S=5/2$) and the easy-plane type of anisotropy.
The following values of the constants in the Hamiltonian
(\ref{H}) were obtained by fitting the measured
magnon frequencies to the classical spin-wave dispersion
\cite{Eibshutz,AZ}: $J=214$~GHz, $J'=0.5$~GHz, $D=1.95$~GHz, and
$g=2$.
It should be mentioned here that the magnon spectrum is also
subjected to quantum renormalizations. If they are not too large the
functional dependences of the magnon frequencies are expected to
preserve the classical form, which involve, however, renormalized
values $\tilde{J}$, $\tilde{J'}$, and $\tilde{D}$. Major
renormalizations in this case are found to be
\cite{Ren}
\begin{equation}
\label{renorm}
\tilde{J}=J\left(1+\frac{\pi -2}{2\pi S} \right), \quad
\tilde{D}=D\left(1-\frac{1}{2S}\right).
\end{equation}
For $S=5/2$ these corrections are not so important
and one can use the above
set of constants to fit the experiment.
With the help of (\ref{<S>3}) we obtain the average value of spin
$S_{\rm av} = 1.82$, which is not far from the experimental
value $S^{\rm exp}_{\rm av} = 1.65$ \cite{Eibshutz}. However, the
notable discrepancy about $10\%$ is quite surprising. It cannot be
explained by the renormalization of constants (\ref{renorm}) and shows
the limits of the harmonic approximation to describe a large spin
reduction even for $S=5/2$. Next we calculate the magnetization curve
for ${\bf H}\parallel\hat{\bf z}$ using the expression (\ref{M3}).
The result is presented in Fig.~2 (solid line) together with
the experimental data \cite{AZ}. Theoretical curve coincides nicely
with the experiment reproducing the unusually low value of the
susceptibility and a small upturn curvature of the magnetization. Better
agreement between the theory and experiment for the uniform
magnetization than for the average spin value is clearly expected
for the quasi one-dimensional
antiferromagnet with a large easy-plane anisotropy $D>J'$.
In the one-dimensional limit ($J'=0$) the average spin value
calculated in the linear spin-wave theory diverges showing its
inapplicability, while the magnetization (\ref{M3}) remains to
be finite.  For ${\bf H}\perp\hat{\bf z}$ we calculate only
the slope of the magnetization curve using the expression
(\ref{chi-par}) for the susceptibility. The resultant curve (dashed
line in Fig.~2) again coincides well with the experiment reproducing
the quantum reduction of the susceptibility. At higher fields
for this orientation of $\bf H$ the
distortions of the triangular structure become important which
requires
to take into account the umklapp terms. At $H>H_c$ the spin structure
transforms into a collinear one \cite{AZ}.

Next we consider CsNiCl$_3$ which has the easy-axis type of anisotropy
and much more pronounced quantum effects due to $S=1$. In order to
calculate the
magnetization for ${\bf H}\perp\hat{\bf z}$ we use the
expressions (\ref{M3}) with $D_{\rm eff}=|D|/2$. For
${\bf H}\parallel\hat{\bf z}$ the magnetization is given
by (\ref{Mxz}) below the
spin-flop transition and by (\ref{M3}) but with negative $D$
above $H_{\rm sf}$. In the case of CsNiCl$_3$ there is a serious
problem to determine values of the constants in the Hamiltonian
(\ref{H}) from the spin-wave dispersions since
the classical expression (22) does not
reproduce the dependences observed on the experiment \cite{Buyers}.
The discrepancy is small for the magnon dispersion
along chains, and
one can estimate from the zone boundary energy
$\tilde{J}=345$~GHz \cite{Buyers}.
With this value of $\tilde{J}$ the fit of
the low-frequency magnetic resonance data and the spin-flop field
value to the classical expressions yield $\tilde{J'}=8$~GHz and
$\tilde{D}=-0.6$~GHz \cite{ZPC}. If we calculate
the magnetization curves using
this
set of parameters, our results show rather poor
agreement with the experimental data. Large
part of the error comes from the use of the renormalized
values for $J$, $J'$, and $D$ instead of the bare Hamiltonian
constants.  For $S=1$ the renormalizations are
quite important:  according to (\ref{renorm}) the bare value of $J$ in
this case is about $18\%$ smaller than $\tilde{J}$, while $D$ is
twice larger than $\tilde{D}$.

On the other hand, the bare value of
the exchange constant $J$ can be determined from the spin-flip field
$H_s$ which is not renormalized due to the absence
of quantum fluctuations
in the saturated phase at $H>H_s$.
Recent high-field magnetization measurements \cite{Goto}
gave $J=275$~GHz and $g=2.15$. We used these values together with
$J'=8$~GHz and $D=-1.6$~GHz to calculate
theoretical curves which are shown in
Fig.~3. The choice of the bare anisotropy constant is rather
arbitrary
since it cannot be calculated from the spin-flop field: for such
large value of the ratio $J/J'$ Eq.~(\ref{Hsf})
fails due to the negative sign of the right
hand side.
Instead, we have chosen the value of $D$ that gives reasonable
description of the difference of magnetizations
for the two perpendicular orientations of magnetic field at
$H>H_{\rm sf}$.  Note, that this difference is
connected with a zero-point contribution to the anisotropy energy,
analogous to the case of CsMnBr$_3$ above $H_c$ \cite{AZ,AP}. Our
calculations with the above constants presented in Fig.~3
give adequate values of the
susceptibilities but distinct difference in the field dependence
of the magnetization.
We note that the disagreement between the classical spin-flop
field depicted in the Figure and the
experimental value does not make any sense since both the classical
$\sqrt{16DJS^2}$ and the renormalized (\ref{Hsf}) expressions
for $H_{\rm sf}$ are inapplicable for
CsNiCl$_3$. This fact together with a poor agreement of the
nonlinear
growth of the magnetization point to the insufficiency of the
harmonic
approximation and need to develop a more accurate treatment including
higher-order terms in $1/S$. Finally, the
sublattice magnetization calculated neglecting the single-ion
anisotropy is $S_{\rm av} = 0.62$ compared with
the experimental value $S^{\rm exp}_{\rm av} = 0.52$.

\section{Conclusions}

We calculated the leading contribution of zero-point fluctuations to
the static properties of a quasi one-dimensional antiferromagnetic
helimagnet. Use of the symmetry of the spin structure
for $D>0$, ${\bf H}\parallel\hat{\bf z}$ and perturbative treatment
of the anisotropy for $D<0$ allowed
us to develop a spin-wave theory with only one kind of boson
operators in the paramagnetic Brillouin zone
and to obtain simple analytic expressions for quantum effect.
Note, that the traditional approach in the framework of
the multisublattice model leads to large mathematical complications
and does not yield analytical results \cite{Watabe}, though
it should be applied in the case of a strong Ising-like anisotropy.

Comparison with the experiment show that theory give good
qualitative description of all specific features observed in the real
quasi one-dimensional antiferromagnets. They include: strong reduction
of the ordered moments and the susceptibility, nonlinear growth of the
transverse magnetization and its anisotropy in the easy-axis case
above
$H_{\rm s-f}$. Accurate quantitative comparison between the theory and
experiment is difficult and give ambiguous results since the bare
constants involved in the Hamiltonian (1) are not known.

Zero-point fluctuations also affect in a different way
different magnetic characteristics. The calculated correction to
the spin-flop field makes its value imaginary in the first
order in $1/S$ for $S=1$ what signals about importance of
higher-order corrections. Instead, the results of the harmonic
approximation give quite good
description of the magnetization curve at high fields even for
a single chain. Corresponding expression (\ref{M4})
fails only in the close neighborhood of the spin-flip
field $H_s$ where it should be corrected via account of the magnon
interactions.

\acknowledgements

M.\,E.\,Zh.\ thanks the Japan Society for the Promotion
of Science for partial financial support on this work.
The work of I.\,A.\,Z. was supported by the Grant M3K000
of the International Science
Foundation, Moscow and by the INTAS Grant N93-1107.

\begin{figure}
\caption{
Magnetization curve for a spin-1 antiferromagnetic chain.
Squares represent the numerical results of
Ref.~[5]. Solid line is the renormalized spin-wave theory
prediction (13), dashed line is the magnetization asymptotics
near the saturation field [3].
}\end{figure}

\begin{figure}
\caption{
Magnetization curves in CsMnBr$_3$. Closed and open circles
are experimental data per mole for ${\bf H}\perp\hat{\bf z}$ and
${\bf H}\parallel\hat{\bf z}$, respectively [14].
Solid line is calculated using
Eq.~(10) with the parameters $J=214$GHz, $J'=0.5$GHz and
$D=1.95$GHz; dashed line is the low-field asymptotics
with the same set of parameters.
}\end{figure}

\vskip 2mm

\begin{figure}
\caption{
Magnetization curves in CsNiCl$_3$. Closed and open circles
are experimental data per mole for ${\bf H}\perp\hat{\bf z}$ and
${\bf H}\parallel\hat{\bf z}$, respectively [15].
Solid lines are the renormalized spin-wave
theory predictions with the parameters $J=275$GHz, $J'=8$GHz and
$D=-1.6$GHz.
}\end{figure}

\end{document}